\documentclass[aps, prb, twocolumn, showpacs, superscriptaddress]{revtex4-1}
\usepackage{graphicx,amsmath,amssymb}
\usepackage[usenames]{color}
\usepackage{indentfirst}
\usepackage{float}
\usepackage[T1]{fontenc}
\usepackage[colorlinks=true, citecolor=blue, urlcolor=blue, linkcolor=blue ]{hyperref}
\hypersetup{breaklinks=true}
\begin{document}
\title{Resistivity minimum emerges in Anderson impurity model modified with Sachdev-Ye-Kitaev interaction}
\author{Lan Zhang}
\affiliation{School of Physical Science and Technology $\&$ Key Laboratory for Magnetism and Magnetic Materials of the MoE, Lanzhou University, Lanzhou 730000, China}
\author{Yin Zhong}
\email{zhongy@lzu.edu.cn}
\affiliation{School of Physical Science and Technology $\&$ Key Laboratory for Magnetism and Magnetic Materials of the MoE, Lanzhou University, Lanzhou 730000, China}
\author{Hong-Gang Luo}
\email{luohg@lzu.edu.cn}
\affiliation{School of Physical Science and Technology $\&$ Key Laboratory for Magnetism and Magnetic Materials of the MoE, Lanzhou University, Lanzhou 730000, China}
\affiliation{Beijing Computational Science Research Center, Beijing 100084, China}
\begin{abstract}
We investigate a modified Anderson model at the large-$N$ limit, where Coulomb interaction is replaced by Sachdev-Ye-Kitaev random interaction. The resistivity of conduction electron $\rho_{c}$ has a minimum value around temperature $\widetilde{T}_{K}$, which is similar to the Kondo system, but the impurity electron's density of state $A_{d}(\omega)$ elucidates no sharp-peak like Kondo resonance around the Fermi surface. The impurity electron's entropy $S_{d}$ and specific heat capacity $C_{\textrm{v}}$ illustrate a crossover from Fermi liquid to the non-Fermi liquid. The system is a non-Fermi liquid at temperature $T^{\star}<T<\widetilde{T}_{K}$, a Fermi liquid for $T<T^{\star}$, and becomes a Fermi gas if $T>\widetilde{T}_{K}$. The non-Fermi liquid at intermediate-$T$ regime does not occur in standard Anderson model. With renormalization group analysis, we elucidate a crossover from Fermi liquid to the non-Fermi liquid, coinciding with transport and thermodynamics. The resistivity minimum and the Kondo resonance are two characteristics of Kondo effect. However, the resistivity minimum emerges in our model when the system behaves as a NFL rather than FL, and the impurity electron's density of state without the Kondo resonance.
\end{abstract}
\maketitle

\section{INTRODUCTION}
The Anderson or Kondo model is at the heart position to understand Kondo physics and heavy fermion compounds.\cite{Hewson1997,Coleman2015,Coleman2007} The first microscopic model for magnetic moments formation in metals is Anderson model, where local moments form once Coulomb interaction between d-electrons becomes large.\cite{PhysRev.124.41} Kondo model is derived from Anderson model via the Schrieffer-Wolff transformation,\cite{PhysRev.149.491} and it demonstrates that resistivity ultimately rises as temperature is lowered, and it connects to the conduction electron's resistivity minimum, which is one of characteristics of the Kondo effect.\cite{ProgTheorPhys.32.37} These two models both have Kondo resonance, i.e. a sharp-peak of the impurity electron's spectral function at Fermi surface, which is the most manifestation when the system appears the Kondo screen to decrease the local moments.\cite{Coleman2015} The evolution from localized magnetic moment state to the non-magnetic state, i.e. from Landau Fermi liquid to the localized Landau Fermi liquid, is a crossover among some rare-earth alloy and actinide compounds.\cite{Hewson1997,Coleman2015}

The Landau Fermi liquid theory has been the workhorse of the physics of interacting electrons for over $60$ years.\cite{Physics3.70} However, some heavy fermion quantum critical compounds such as CeCu$_{6-x}$Au$_{x}$, YbRh$_{2}$Si$_{2}$ and $\beta$-YbAlB$_{4}$ display the non-Fermi liquid (NFL) behavior that the transport property and specific heat capacity are deviated from Fermi liquid (FL).\cite{schroder2000onset,custers2003,matsumoto2011quantum} This attracts much attention on the NFL behavior, and many theories are formulated to interpret this phenomenon.\cite{vcubrovic2009string,si2001,RevModPhys.78.17,PhysRevLett.94.066402,PhysRevLett.90.216403,PhysRevB.69.035111,Yang8398} The lack of controlled theoretical techniques hinders the understanding of the strong electron correlation in NFL, until the invention of Sachdev-Ye-Kitaev (SYK) model.\cite{PhysRevX.8.031024,PhysRevB.97.241106,PhysRevB.95.205105,PhysRevB.97.155117}

SYK model is a quantum many-body model with random all-to-all interactions for fermions, which was studied in the 1990s and later as models for novel NFL or spin-glass states.\cite{PhysRevLett.70.3339,PhysRevB.59.5341,PhysRevLett.85.840,PhysRevB.63.134406,PhysRevLett.90.217202,Kitaev1,Kitaev2} It provides a solvable example in zero dimension and has been extended to higher dimensions.\cite{Berkooz2017,Gu2017,PhysRevLett.105.151602,Polchinski2016,PhysRevX.5.041025,PhysRevD.94.106002,PhysRevX.7.031006} In recent years, many exotic physical phenomena have been found in SYK models, e.g. supersymmetry,\cite{PhysRevD.95.026009} quantum chaos,\cite{PhysRevB.96.060301,Gu2017,PhysRevB.96.205138,Hosur2016} many-body localization,\cite{PhysRevLett.119.206602,PhysRevB.95.115150} strongly correlated metal,\cite{PhysRevLett.119.216601,PhysRevX.8.021049} and quantum phase transition.\cite{PhysRevB.95.134302,PhysRevB.97.155117,PhysRevB.95.205105,PhysRevLett.119.207603,PhysRevB.97.205141,PhysRevB.98.165135}

In a recent work with aim to provide a solvable model for heavy fermion system, the standard periodic Anderson model is modified with the SYK random interaction. They find a low-temperature FL and more interestingly a NFL solution at elevated temperature, and the rising of resistivity at high temperature is claimed to result from the single SYK quantum impurity.\cite{JPC2018} Considering the distinction between impurity and lattice model, in this work, we study the SYK quantum impurity problem, modeled by Anderson model with SYK random interaction. (We call it Sachdev-Ye-Kitaev Anderson model (SYKAM).)

Under the large-N limit, the qualitative analysis of the conduction electron resistivity $\rho_{c}$ elucidates that the SYKAM behaves as FL when temperature $T<T^{\star}$ and is a NFL at temperature $T>T^{\star}$, where $T^{\star}$ is a scaling temperature. More quantitative calculation shows that $\rho_{c}$ exists a minimum at temperature $\widetilde{T}_{K}$, demonstrating SYKAM behaves as FL when $T<T^{\star}$ and NFL at low temperature $T^{\star}<T<\widetilde{T}_{K}$. At high temperature $T>\widetilde{T}_{K}$, a free Fermi gas (FG) forms in SYKAM. It is emphasized that NFL does not display in standard Anderson model. From impurity electron's entropy $S_{d}$ and specific heat capacity $C_{\textrm{v}}$, a crossover, not a phase transition, exists between FL and NFL. This is confirmed by a renormalization group (RG) analysis, whose flow equation is similar to Kondo problem.\cite{Hewson1997}

What is more, our system does not form the local moment, because the SYK random interaction does not provide localized interaction to the onsite different impurity spin states. The hybridization between conduction and impurity electrons contributes to extending the localized impurity electron's density of state (DOS) to the Lorentz-like lineshape, and the impurity electron has the scattering with conduction electron sea, but its DOS shows that there is no sharp-peak at the Fermi surface like Kondo resonance. \cite{Hewson1997,Coleman2015} Since the resistivity minimum and the Kondo resonance are two characteristics of Kondo effect, in which the conduction electron screens the impurity electron's local moments.\cite{Hewson1997,Coleman2015} The resistivity minimum emerges in our model when the system behaves as a NFL rather than FL, and the DOS of the impurity electron without the Kondo resonance.

The paper is organized as follows. In Sec.~\ref{sec:1}, we introduce SYKAM and derive self-energy and the impurity electron Green's function. In Sec.~\ref{sec:4}, we present the conduction electron resistivity $\rho_{c}$. In Sec.~\ref{sec:5}, we compute thermodynamics, i.e. impurity electron's entropy $S_{d}$ and its specific heat capacity $C_{\textrm{v}}$. In Sec.~\ref{sec:6}, we apply RG theory to analyse above results. Finally, Sec.~\ref{sec:7} is devoted to a brief conclusion and perspective.
\section{MODEL AND METHOD}\label{sec:1}
\begin{figure}
  \centering
  \includegraphics[width=0.8\columnwidth]{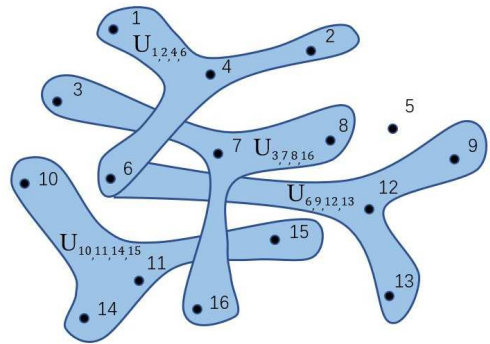}\\
  \caption{The SYK random interaction $U_{ijml}$ between different impurity electron pseudospin states $1, 2, ..., 16, ..., N$.}\label{fig:p3}
\end{figure}

The Hamiltonian of SYKAM can be written as
\begin{eqnarray}\label{eq:1}
H=&&\sum_{kj}\varepsilon_{k}\hat{c}^{\dag}_{kj}\hat{c}_{kj}+V\sum_{kj}\left(\hat{c}^{\dag}_{kj}\hat{d}_{j}+\hat{d}^{\dag}_{j}\hat{c}_{kj}\right)\nonumber\\
&&+E_{d}\sum_{j}\hat{d}^{\dag}_{j}\hat{d}_{j}+\frac{1}{(2N)^{3/2}}\sum_{ijml}U_{ijml}\hat{d}^{\dag}_{i}\hat{d}^{\dag}_{j}\hat{d}_{m}\hat{d}_{l}.
\end{eqnarray}
Here $\hat{c}^{\dag}_{kj}$ ($\hat{c}_{kj}$) and $\hat{d}^{\dag}_{j}$ ($\hat{d}_{j}$) denote the creation (annihilation) operator of conduction and impurity electrons with pseudospin $j=1,2,...,N$, respectively. In the first line of Eq.~(\ref{eq:1}), conduction electrons have energy dispersion $\varepsilon_{k}$ and hybridize with impurity electron with strength $V$. The impurity electron has degenerated energy level $E_{d}$. There exists SYK-like random all-to-all interaction $U_{ijml}$ between different pseudospin states of impurity electron, and $U_{ijml}$ has standard Gaussian random distribution with zero average $\overline{U}_{ijml}=0$ and variance $\overline{U^{2}}_{ijml}=U^{2}$ as shown in Fig. \ref{fig:p3}.
\begin{figure}
  \centering
  \includegraphics[width=0.8\columnwidth]{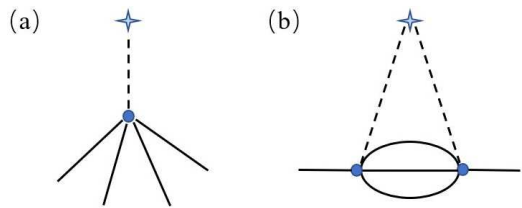}\\
  \caption{The leading impurity electron self-energy Feynman diagrams for SYKAM under the large-N limit. The left panel (a) is the bare interaction vertex before average over random interaction $U_{ijml}$. The right panel (b) illustrates the self-energy after average. }\label{fig:p4}
\end{figure}

Under the large-N limit, we can obtain the leading order Feynman diagrams as shown in Fig.~\ref{fig:p4}. The resulting conduction electrons Green's function $G_{c}(k,i\omega_{n})$ and the impurity electron Green's function $G_{d}(i\omega_{n})$ are given by
\begin{equation}
G_{c}(k,i\omega_{n})=\frac{1}{i\omega_{n}-\varepsilon_{k}-V^{2}G_{d}(i\omega_{n})},
\end{equation}
\begin{equation}\label{eq:15}
G_{d}(i\omega_{n})=\frac{1}{i\omega_{n}-E_{d}-\sum_{k}\frac{V^{2}}{i\omega_{n}-\varepsilon_{k}}-\Sigma(i\omega_{n})},
\end{equation}
where $\omega_{n}=(2n+1)\pi T$ denotes the fermionic Matsubra frequency with $n=0,\pm1,\pm2,...,\pm\infty$. The impurity electron imaginary-time self-energy is
\begin{eqnarray}\label{eq:5}
&&\Sigma(\tau)=U^{2}\left[G_{d}(\tau)\right]^{2}G_{d}(-\tau),
\end{eqnarray}
where $G_{d}(\tau)$ is the imaginary-time impurity electron Green's function.

Consequently, Green's functions $G_{d}(i\omega_{n})$ and $G_{c}(k,i\omega_{n})$ can be found by solving Eqs.~(\ref{eq:15}) and (\ref{eq:5}) self-consistently. In order to get the analytic results of Eqs.~(\ref{eq:15}) and (\ref{eq:5}), we consider two limiting cases as follows: the weak coupling limit $|i\omega_{n}|\gg|\Sigma(i\omega_{n})|$ and the strong coupling limit $|\Sigma(i\omega_{n})|\gg |i\omega_{n}|$.
\subsection{WEAK COUPLING LIMIT}\label{sec:2}
In the weak coupling limit $|i\omega_{n}|\gg|\Sigma(i\omega_{n})|$, we can compute the impurity electron Green's function via the perturbation theory of the random interaction term. The free impurity electron Green's function (without the SYK interaction) is
\begin{equation}
G^{0}_{d}(i\omega_{n})=\frac{1}{i\omega_{n}-E_{d}-\sum_{k}\frac{V^{2}}{i\omega_{n}-\varepsilon_{k}}}.
\end{equation}
For simplicity, the DOS of conduction electrons is assumed to be $N(\varepsilon)=\frac{1}{2D}\theta(D-|\varepsilon|)$ with $\theta(x)$ being a step function, and $2D$ is the band-width of conduction electrons. Under above assumption, the hybridization contribution to the impurity electron is
\begin{eqnarray}
\sum_{k}\frac{V^{2}}{i\omega_{n}-\varepsilon_{k}}=-V^{2}N(0)\ln\left[\frac{D-i\omega_{n}}{-D-i\omega_{n}}\right].
\end{eqnarray}
When $D\gg|\omega_{n}|$, we get
\begin{equation}
G^{0}_{d}(i\omega_{n})=\frac{1}{i\omega_{n}-E_{d}+i\Delta\textrm{sgn}(i\omega_{n})},
\end{equation}
where $\Delta=\pi V^{2}N(0)$, and $N(0)=\frac{1}{2D}$ is the DOS of the conduction electron at Fermi energy.\cite{Li2002} Hence, the impurity electron DOS is
\begin{equation}\label{eq:3}
A^{0}_{d}(\omega)=\frac{1}{\pi}\frac{\Delta}{(\omega-E_{d})^{2}+\Delta^{2}}.
\end{equation}

Thus, the impurity electron self-energy Eq.~(\ref{eq:5}) becomes
\begin{eqnarray}
\Sigma(i\omega_{n})&&=U^{2}\int d\omega_{1}\int d\omega_{2}\int d\omega_{3}\bigg[A^{0}_{d}(\omega_{1})A^{0}_{d}(\omega_{2})A^{0}_{d}(\omega_{3})\nonumber\\
&&\quad\left.\frac{f(\omega_{1})f(-\omega_{2})f(-\omega_{3})+f(-\omega_{1})f(\omega_{2})f(\omega_{3})}{i\omega_{n}+\omega_{1}-\omega_{2}-\omega_{3}}\right],\nonumber\\
&&
\end{eqnarray}
where $f(x)$ is the Fermi-Dirac distribution function.

At zero temperature, we use the analytic continuation $i\omega_{n}\rightarrow\omega+i\delta$ to get the zero temperature impurity electron self-energy
\begin{eqnarray}
\Sigma(\omega)=&&U^{2}\int d\omega_{1}\int d\omega_{2}\int d\omega_{3}\bigg[A^{0}_{d}(\omega_{1})A^{0}_{d}(\omega_{2})A^{0}_{d}(\omega_{3})\nonumber\\
&&\left.\frac{\theta(-\omega_{1})\theta(\omega_{2})\theta(\omega_{3})+\theta(\omega_{1})\theta(-\omega_{2})\theta(-\omega_{3})}{\omega+i\delta+\omega_{1}-\omega_{2}-\omega_{3}}\right],\nonumber\\
&&
\end{eqnarray}
where $\delta$ denotes the infinitesimal positive parameter. The imaginary part of the impurity self-energy $\Sigma(\omega)$ is
\begin{eqnarray}\label{eq:13}
\rm{Im}\Sigma(\omega)=&&-U^{2}\pi\left\{\int^{\infty}_{0}d\omega_{2}\int^{\omega-\omega_{2}}_{0}d\omega_{3}\Big[\theta(\omega-\omega_{2})\right.\nonumber\\
&&A^{0}_{d}(\omega_{2}+\omega_{3}-\omega)A^{0}_{d}(\omega_{2})A^{0}_{d}(\omega_{3})\Big]\nonumber\\
&&+\int^{0}_{-\infty}d\omega_{2}\int^{0}_{\omega-\omega_{2}}d\omega_{3}\Big[\theta(\omega_{2}-\omega)\nonumber\\
&&A^{0}_{d}(\omega_{2}+\omega_{3}-\omega)A^{0}_{d}(\omega_{2})A^{0}_{d}(\omega_{3})\Big]\bigg\}.
\end{eqnarray}
To proceed, we set $E_{d}=0$ and consider the low-energy limit with $|\omega|\ll\Delta$, and the impurity electron DOS of Eq. (\ref{eq:3}) is
\begin{eqnarray}
A^{0}_{d}(\omega)&&=\frac{1}{\pi\Delta}\left[1-\frac{\omega^{2}}{\Delta^{2}}+\emph{O}\left(\frac{\omega^{4}}{\Delta^{4}}\right)\right]\approx\frac{1}{\pi\Delta}.
\end{eqnarray}
Thus, Eq. (\ref{eq:13}) is approximated to be
\begin{eqnarray}
\rm{Im}\Sigma(\omega)=-\frac{U^{2}\pi}{2(\pi\Delta)^{3}}\omega^{2}+\emph{O}(\omega^{4})\propto\omega^{2},
\end{eqnarray}
which is an essential feature of FL.\cite{AGD1964} Via Kramers-Kronig relation, its real part is $\rm{Re}\Sigma(\omega)\propto\omega$.\cite{Coleman2015} To beyond perturbation theory analysis, we assume that the impurity electron Green's function has the following FL-like form
\begin{equation}\label{eq:4}
G_{d}(\omega)=\frac{Z}{\omega+iB\Delta}+\mathrm{incoherent},
\end{equation}
where $Z$ is the quasiparticle weight, and $B$ is an unknown parameter. Then, the imaginary part of the impurity electron self-energy reads
\begin{equation}
\rm{Im}\Sigma(\omega)\simeq-\left(\frac{Z}{B}\right)^{3}\frac{U^{2}\pi}{2(\pi\Delta)^{3}}\omega^{2}.
\end{equation}
and its real part is
\begin{equation}
\rm{Re}\Sigma(\omega)\simeq-\left(\frac{Z}{B}\right)^{3}\left(\frac{U}{\pi\Delta}\right)^{2}\omega.
\end{equation}
Therefore, we obtain
\begin{eqnarray}
G_{d}(\omega)\simeq\frac{1}{\omega+i\Delta+\left(\frac{Z}{B}\right)^{3}\left(\frac{U}{\pi\Delta}\right)^{2}\omega}.
\end{eqnarray}
Comparing with Eq.~(\ref{eq:4}), we find the quasiparticle weight $Z=\frac{1}{1+\left(\frac{U}{\pi\Delta}\right)^2}$ and $B=Z$. We conclude that the system behaves like a local FL, which is similar to the ground state of standard Anderson impurity model.\cite{Hewson1997}
\subsection{STRONG COUPLING LIMIT}\label{sec:3}
In the strong coupling limit $|\Sigma(i\omega_{n})|\gg |i\omega_{n}|$, we can neglect the bare $i\omega_{n}$ term, so the impurity electron Green's function is
\begin{equation}\label{eq:7}
G_{d}(i\omega_{n})=-\frac{1}{\Sigma(i\omega_{n})}.
\end{equation}
Due to Eq. (\ref{eq:5}), we can formulate the Matsubara Green's function at the zero temperature limit as
\begin{equation}
G_{d}(\tau)=\frac{b}{|\tau|^{2x}}\rm{sgn}(\tau),
\end{equation}
where $x$ and $b$ are unknown parameters. With the dimension analysis, we have $x=\frac{1}{4}$ and a straightforward calculation gives
$b=\frac{1}{\sqrt[4]{4U^{2}\pi}}$.\cite{PhysRevD.94.106002} So, we have

\begin{equation}
G_{d}(\tau)=\frac{1}{\sqrt[4]{4U^{2}\pi}}\frac{1}{|\tau|^{\frac{1}{2}}}\rm{sgn}(\tau).
\end{equation}
Under the Fourier transformation and analytic continuation,
\begin{eqnarray}\label{eq:8}
&&G_{d}(i\omega_{n})=\int^{\beta}_{0}d\tau e^{-i\omega_{n}\tau}\frac{\pi^{-1/4}\rm{sgn(\tau)}}{\sqrt{2U|\tau|}}=\frac{\pi^{-1/4}}{\sqrt{2U}}\frac{\sqrt{\pi}}{\sqrt{-i\omega_{n}}},\nonumber\\
&&G_{d}(\omega)=\frac{\pi^{-1/4}}{\sqrt{2U}}\frac{\sqrt{\pi}}{\sqrt{|\omega|}}\theta(-\omega)-i\frac{\pi^{-1/4}}{\sqrt{2U}}\frac{\sqrt{\pi}}{\sqrt{|\omega|}}\theta(\omega),
\end{eqnarray}
and the quasiparticle weight is found to be vanished as
\begin{eqnarray}
Z=\left.\frac{1}{1-\partial_{\omega}\rm{Re\Sigma}}\right|_{\omega\rightarrow 0}\propto\left.\sqrt{\frac{|\omega|}{U}}\right|_{\omega\rightarrow0}\rightarrow 0.
\end{eqnarray}
Therefore, the system is a NFL without any quasiparticle.
\section{TRANSPORT PROPERTIES}\label{sec:5}
Before immersing into the calculation of transport quantities, we inspect the behavior of single-particle Green's functions. Firstly, by comparing the hybridization term $\Delta$ with strong coupling self-energy, there exists a characteristic energy scale
\begin{equation}\label{eq:20}
E^{\star}\sim\frac{\Delta^{2}}{U}.
\end{equation}
At $|\omega|\gg E^{\star}$, the impurity electron Green's function is dominated by the self-energy term and has the NFL feature. When $|\omega|\ll E^{\star}$, it shows FL-like behavior, i.e. ($T=0$)
\begin{eqnarray*}
G_{d}(\omega)=\left\{
                    \begin{array}{cc}
                    \frac{1}{\omega+i\Delta+\left(\frac{U}{\pi\Delta}\right)^{2}\omega
                    +i\frac{U^{2}\pi}{2(\pi\Delta)^{3}}\omega^{2}}, ~~|\omega|\ll E^{\star}.\\
                    ~~~~~~~~~\frac{\pi^{1/4}}{\sqrt{2U}}\frac{1}{\sqrt{|\omega|}},~~~~~~~~~~~|\omega|\gg E^{\star}.
                    \end{array}
              \right.
\end{eqnarray*}
Also, the conduction electron has its Green's function as
\begin{eqnarray*}
G_{c}(k,\omega)=\left\{
                    \begin{array}{cc}
                    \frac{1}{\omega-\varepsilon_{k}-\frac{V^{2}}{\omega+i\Delta+\left(\frac{U}{\pi\Delta}\right)^{2}\omega
                    +i\frac{U^{2}\pi}{2(\pi\Delta)^{3}}\omega^{2}}}, ~~|\omega|\gg E^{\star}. \\
                    ~~~~~~~\frac{1}{\omega-\varepsilon_{k}-\frac{V^{2}\pi^{1/4}}{\sqrt{2U}}\frac{1}{\sqrt{|\omega|}}},~~~~~~~~~~|\omega|\ll E^{\star}.
                    \end{array}
              \right.
\end{eqnarray*}
The scattering rate of conduction electron is estimated to be
\begin{eqnarray}\label{eq:9}
\frac{1}{\tau_{c}}\propto-\rm{Im}\Sigma_{c}(\omega)\sim\left\{
                    \begin{array}{cc}
                    ~\frac{V^{2}}{\Delta}, ~~\quad|\omega|\ll E^{\star},\\
                    \frac{V^{2}}{\sqrt{|\omega|}}, \quad|\omega|\gg E^{\star}.
                    \end{array}
    \right.
\end{eqnarray}
Assuming the transport scattering rate is proportional to the above single-particle one $\frac{1}{\tau_{c}}$, the finite temperature resistivity is found to be
\begin{eqnarray}\label{eq:14}
\rho_{c}\propto\frac{1}{\tau_{c}}\propto
\left\{
                    \begin{array}{cc}
                    \frac{V^{2}}{\Delta}-T^{2}, ~~\quad T\ll T^{\star}\\
                    \frac{V^{2}}{\sqrt{U}}T^{-1/2}, \quad T\gg T^{\star}
                    \end{array}
    \right.
\end{eqnarray}

Quantitatively, the static resistivity is in inverse proportion to the static limit of optical conductivity. The latter can be obtained via Kubo formula\cite{Coleman2015}
\begin{equation}
\sigma_{xx}(\omega)=\frac{1}{\omega}\Pi^{\prime\prime}(\omega+i\delta),
\end{equation}
and
\begin{equation}
\Pi_{xx}(\omega)=-i\int^{\infty}_{0}e^{i\omega t}\langle[\hat{J}_{x}(t),\hat{J}_{x}(0)]\rangle
\end{equation}
denotes the current-current response function. Here, the $x$-component of current operator $\hat{J}_{x}$ is given by\cite{Czycholl1981}
\begin{equation}
\hat{J}_{x}=\frac{e}{\hbar}\sum_{kj}\frac{\partial\varepsilon_{k}}{\partial k_{x}}\hat{c}^{\dag}_{kj}\hat{c}_{kj}.
\end{equation}
Inserting $G_{c}(k,\omega)$ into $\sigma_{xx}(\omega)$, the static conductivity is
\begin{eqnarray}\label{eq:21}
\sigma_{xx}(\omega=0)=\int d\varepsilon_{1}\left[-\frac{df}{d\varepsilon_{1}}\right]L(\varepsilon_{1}),
\end{eqnarray}
where $L(\varepsilon_{1})=\frac{e^{2}\pi}{\hbar^{2}}\sum_{k\sigma}\left(\frac{\partial\varepsilon_{k}}{\partial k_{x}}\right)^{2}\left(A_{c}(k,\varepsilon_{1})\right)^{2}$, and the spectral function of conduction electrons is $A_{c}(k,\varepsilon_{1})=-\frac{1}{\pi}\rm{Im}G_{c}(k,\varepsilon_{1}+i\delta)$.
\begin{figure}
\includegraphics[width=0.48\columnwidth]{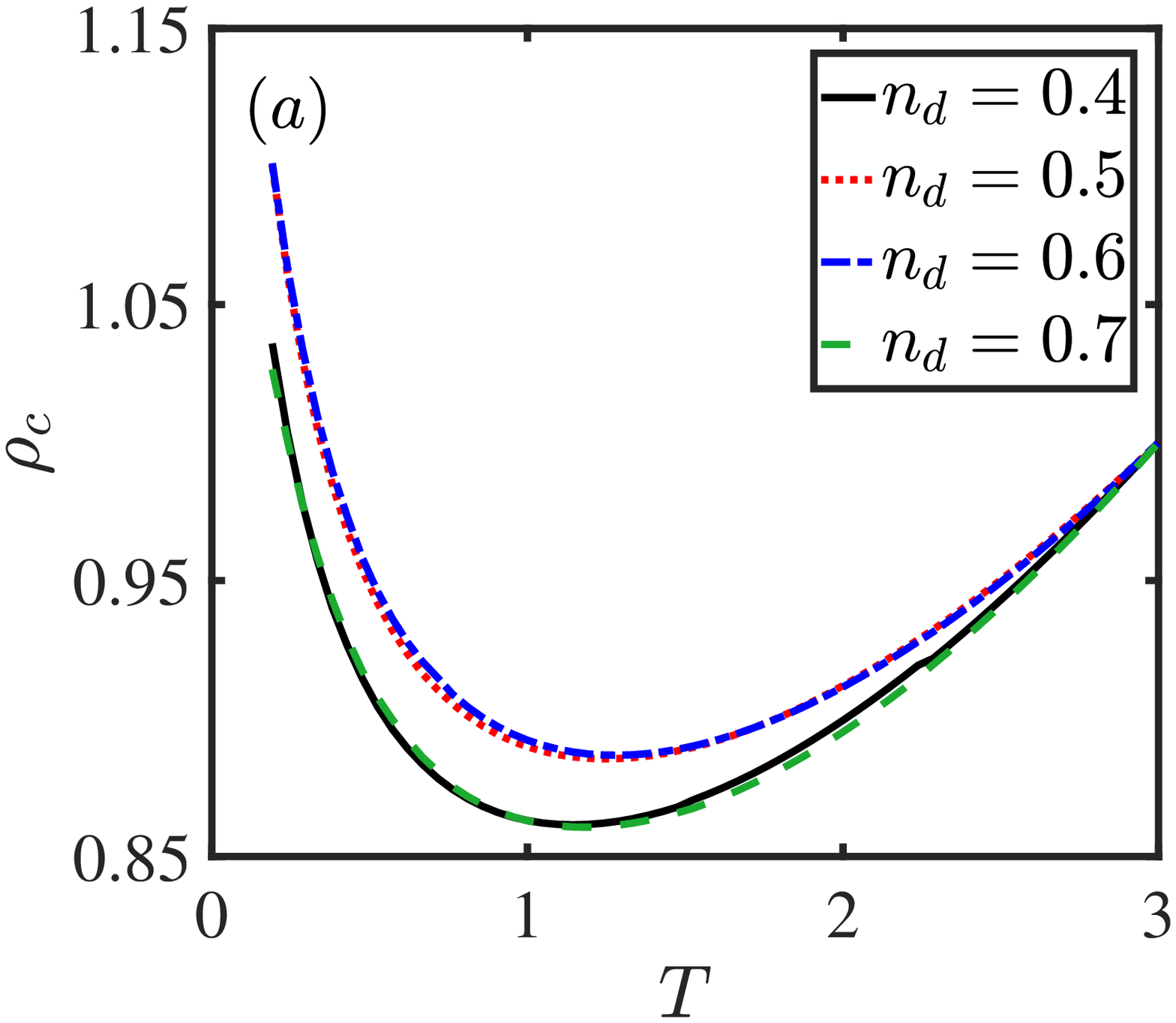}\quad
\includegraphics[width=0.48\columnwidth]{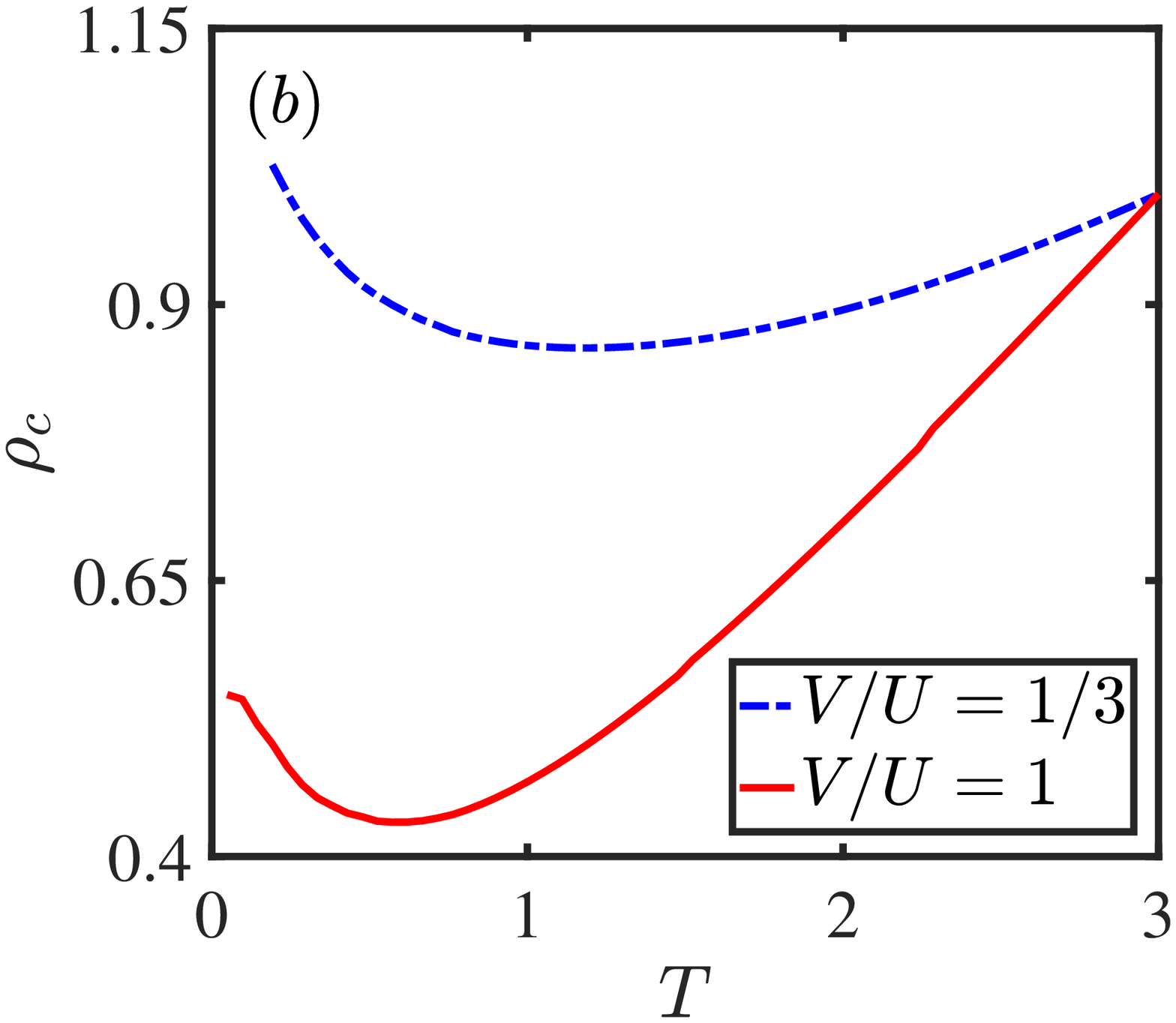}
\caption{The conduction electron resistivity $\rho_{c}$ as a function of temperature $T$. For the different impurity electron concentrations $n_{d}$ as shown in subplot (a),  and the parameters are $V=1, U=3, D=12, E_{d}=-U/2$. In subplot (b), $\rho_{c}$ versus $V/U$ at $n_{d}=0.7$. The blue dashed line is the same as $n_{d}=0.7$ in (a) with $\widetilde{T}_{K}=1.2387$, and the red solid line is $V/U=1/3$ with $\widetilde{T}_{K}=0.3342$, where parameters are $V=1, U=1, D=4, E_{d}=-U/2$. }\label{fig:p2}
\end{figure}
\begin{figure}
  \centering
  \includegraphics[width=\columnwidth]{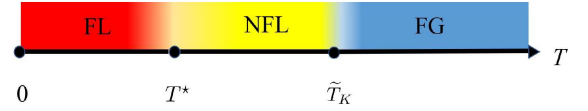}\\
  \caption{The phase transition versus temperature in the SYKAM. Our system has the FL behavior at $T<T^{\star}$ and behaves like NFL at $T^{\star}<T<\widetilde{T}_{K}$, while it is FG at $T>\widetilde{T}_{K}$.}\label{fig:p7}
\end{figure}
\section{THERMODYNAMICS}\label{sec:4}
Thermodynamics of our model is determined by the partition function $\mathcal{Z}$, whose functional integral formalism is
\begin{equation}
\mathcal{Z}=\int\mathcal{D}c^{\dag}\mathcal{D}c\mathcal{D}d^{\dag}\mathcal{D}de^{-\mathcal{S}_{0}-\mathcal{S}_{int}}.
\end{equation}
Here the non-interacting action is
\begin{eqnarray}
\mathcal{S}_{0}=&&\int d\tau\left[\sum_{kj}c^{\dag}_{kj}\left(\partial_{\tau}+\varepsilon_{k}\right)c_{kj}
+V\sum_{kj}c^{\dag}_{kj}d_{j}\right.\nonumber\\
&&\left.+V\sum_{kj}d^{\dag}_{j}c_{kj}+\sum_{j}d^{\dag}_{j}\left(\partial_{\tau}+E_{d}\right)d_{j}\right],
\end{eqnarray}
with imaginary time $\tau\in[0,\beta]$ ($\beta=1/T$) and $c_{kj}$, $d_{j}$ are the anticommuting Grassman fields. The SYK interaction reads
\begin{eqnarray}
&&\mathcal{S}_{int}=-\frac{1}{(2N)^{3/2}}\int d\tau\sum_{ijml}U_{ijml}d^{\dag}_{i}d^{\dag}_{j}d_{m}d_{l}.
\end{eqnarray}
After performing the standard Gaussian random average over each independent $U_{ijml}$ and focusing on one replica realization,\cite{PhysRevLett.119.216601} we obtain
\begin{eqnarray}\label{eq:16}
\mathcal{S}_{int}=&&-\frac{U^{2}}{4N^{3}}\int d\tau\int d\tau^{\prime}\sum_{ijml}\left[d^{\dag}_{i}(\tau)d_{i}(\tau^{\prime})\right.\nonumber\\
&&\left. d^{\dag}_{j}(\tau)d_{j}(\tau^{\prime})d^{\dag}_{m}(\tau^{\prime})d_{m}(\tau)d^{\dag}_{l}(\tau^{\prime})d_{l}(\tau)\right].
\end{eqnarray}

Now we introduce $G_{d}(\tau^{\prime},\tau)=\frac{1}{N}\sum_{j}d^{\dag}_{j}(\tau)d_{j}(\tau^{\prime})$ and a Lagrange multiplier $\Sigma(\tau,\tau^{\prime})$ into the action $\mathcal{S}_{int}$ with adding the following constraint term into the partition function,
\begin{eqnarray}
1&&=\int\mathcal{D}G\delta\left(G_{d}(\tau^{\prime},\tau)-\frac{1}{N}\sum_{j}d^{\dag}_{j}(\tau)d_{j}(\tau^{\prime})\right)\nonumber\\
&&=\int\mathcal{D}\Sigma\int\mathcal{D}Ge^{\int d\tau\int d\tau^{\prime}\Sigma(\tau,\tau^{\prime})\big[NG_{d}(\tau^{\prime},\tau)-\sum_{j}d^{\dag}_{j}(\tau)d_{j}(\tau^{\prime})\big]}.\nonumber\\
&&
\end{eqnarray}
Therefore, we can rewrite $\mathcal{S}_{int}$ as
\begin{eqnarray}
\mathcal{S}_{int}&&=-\frac{U^{2}}{4N^{3}}\int d\tau\int d\tau^{\prime}(G_{d}(\tau^{\prime},\tau))^{2}(G_{d}(\tau,\tau^{\prime}))^{2}\nonumber\\
&&=-\frac{U^{2}}{4N^{3}}\int d\tau\int d\tau^{\prime}|G_{d}(\tau^{\prime},\tau)|^{4},
\end{eqnarray}
where $G_{d}(\tau^{\prime},\tau)=G^{\star}_{d}(\tau,\tau^{\prime})$. After integrating out conduction electrons, the local electron only action is
\begin{eqnarray}
\mathcal{S}_{d}=&&\int d\tau\int d\tau^{\prime}\left\{\sum_{j}d^{\dag}_{j}(\tau)\Bigg[\Sigma(\tau,\tau^{\prime})+\Bigg(\partial_{\tau^{\prime}}+E_{d}\right.\nonumber\\
&&\left.\left.\left.-\sum_{k}V^{2}\left(\partial_{\tau}+\varepsilon_{k}\right)^{-1}\right)\delta(\tau-\tau^{\prime})\right]d_{j}(\tau^{\prime})\right\}\nonumber\\
&&-N\int d\tau\int d\tau^{\prime}\Sigma(\tau,\tau^{\prime})G_{d}(\tau,\tau^{\prime})\nonumber\\
&&-\frac{NU^{2}}{4}\int d\tau\int d\tau^{\prime}|G_{d}(\tau,\tau^{\prime})|^{4},
\end{eqnarray}
where $\mathcal{S}_{d}$ is the effective action for $G$ and $\Sigma$. In the large-N limit, the partition function is dominated by the extremal $\mathcal{S}_{d}$, which leads to Eqs.~(\ref{eq:15}) and (\ref{eq:5}) via the saddle point equations
\begin{eqnarray}
\frac{\delta \mathcal{S}_{d}}{\delta G}=0,~~\frac{\delta \mathcal{S}_{d}}{\delta\Sigma}=0.
\end{eqnarray}
Therefore, the impurity electron contributes a free-energy as
\begin{eqnarray}
\frac{F_{d}}{N}=&&-T\ln2-\frac{3}{4}T\sum_{n}\Sigma(i\omega_{n})G_{d}(i\omega_{n})\nonumber\\
&&+T\sum_{n}\Big[\ln(-\beta G_{d}(i\omega_{n}))-\ln(-\beta i\omega_{n})\Big],
\end{eqnarray}
while the conduction electron has FG result
$\frac{F_{c}}{N}=-T\sum_{k}\ln(1+e^{-\beta\varepsilon_{k}}).$
In this way, the system's total free-energy $F$ can be obtained by
$\frac{F}{N}=\frac{F_{c}}{N}+\frac{F_{d}}{N}$.

Because conduction electrons only contribute trivial FG result, we focus on the thermodynamics of the impurity electron. Therefore, the impurity electron's entropy $S_{d}$ and specific heat capacity $C_{\textrm{v}}$ are given by
\begin{equation}\label{eq:11}
S_{d}=-\frac{\partial F_{d}}{\partial T},~~C_{\textrm{v}}=T\frac{\partial S_{d}}{\partial T}=-T\frac{\partial^2 F_{d}}{\partial T^2}.
\end{equation}
\begin{figure}
\includegraphics[width=0.48\columnwidth]{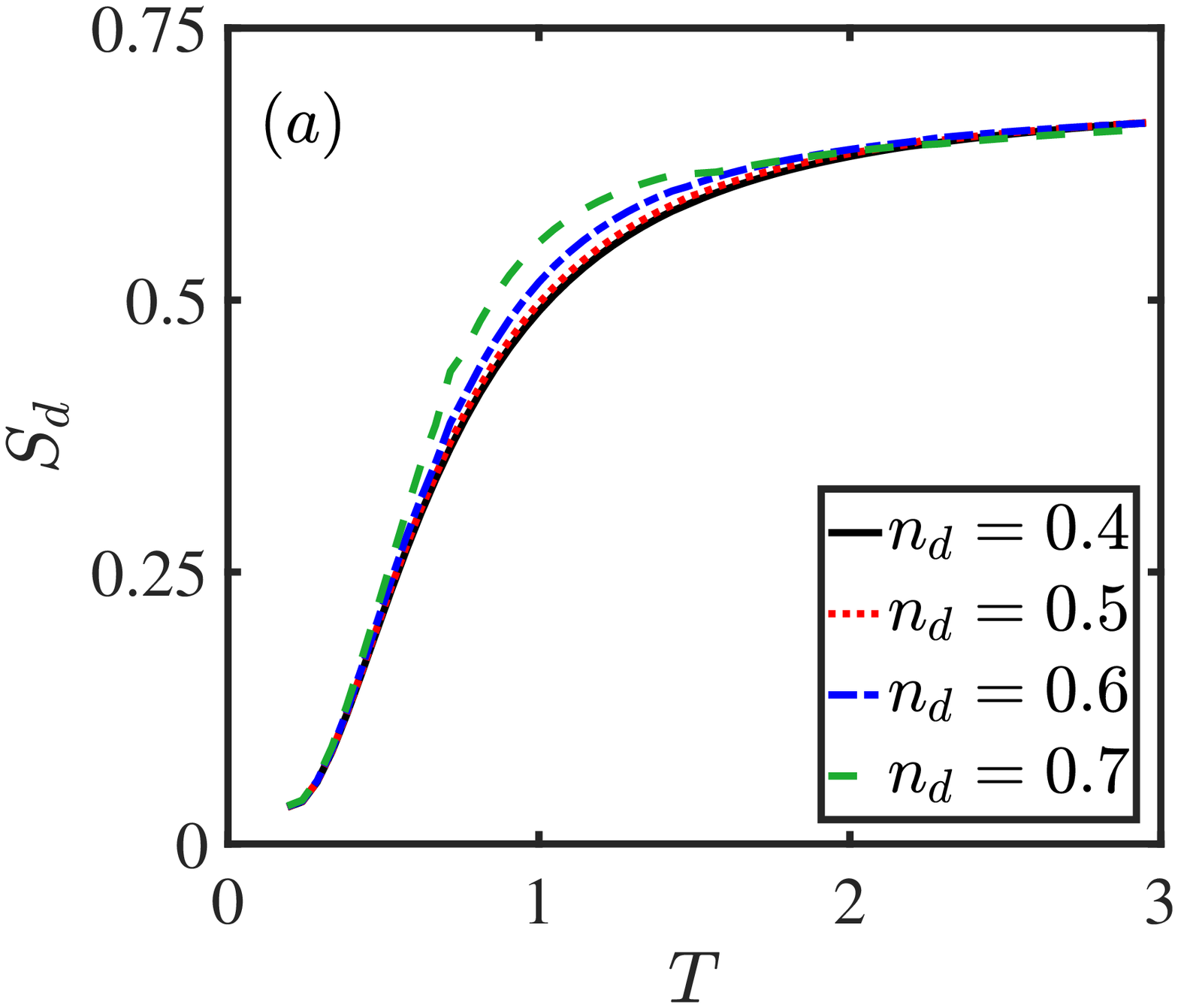}\quad
\includegraphics[width=0.48\columnwidth]{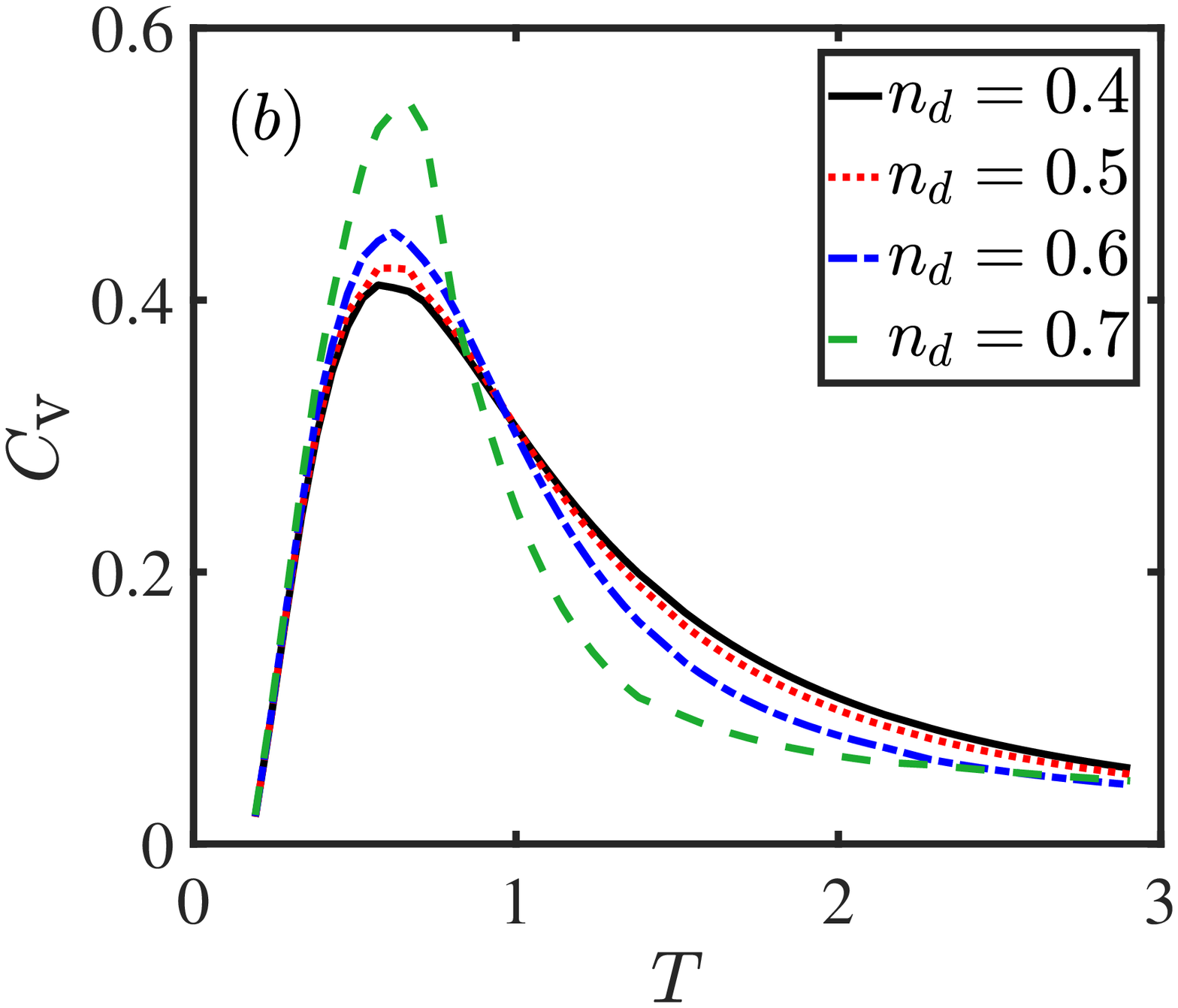}\\
\caption{The thermodynamics of our model as a function of temperature $T$, for different impurity electron concentrations $n_{d}$. $(a)$ the entropy $S_{d}$ and $(b)$ the specific heat capacity $C_{\textrm{v}}$ of the impurity electron. The parameters are $V=1, U=3, D=12, E_{d}=-U/2$. }\label{fig:p1}
\end{figure}
\section{ANALYTICAL ANALYSIS}\label{sec:6}
From the Eq.~(\ref{eq:21}), we plot the conduction electron resistivity $\rho_{c}$ versus $T$, as shown in Fig.~\ref{fig:p2}. $\rho_{c}$ has the minimum at $\widetilde{T}_{K}$, which is similar to Kondo effect.\cite{PhysRev.135.A1041,Coleman2015,Hewson1997} The line of $n_{d}=0.4$ coincides with $n_{d}=0.7$, which is similar to $n_{d}=0.5$ and $n_{d}=0.6$. It means that our system is symmetric about the $n_{d}=0.55$ because of $E_{d}=-U/2$ in Fig.~\ref{fig:p2} (a). In Fig.~\ref{fig:p2} (b), it demonstrates that the line of $V/U=1/3$ with $D=12$ has the higher $\widetilde{T}_{K}$ than $V/U=1$ with $D=4$, where the wide bandwidth $2D$ and the large random interaction $U$ induces a high $\widetilde{T}_{K}$. The red solid line of Fig.~\ref{fig:p2} (b) demonstrates that our system has the FL behavior at $T<T^{\star}$ and behaves like NFL at $T^{\star}<T<\widetilde{T}_{K}$, while it is FG at $T>\widetilde{T}_{K}$ as shown in Fig.~\ref{fig:p7}. According to Eq.~(\ref{eq:20}), $V/U=1$ has the higher $E^{\star}$ and $T^{\star}$ than $V/U=1/3$, thus $E^{\star}$ and $T^{\star}$ of the $V/U=1/3$ are too low to show FL behavior in Fig.~\ref{fig:p2}.

\begin{figure}
  \includegraphics[width=0.48\columnwidth]{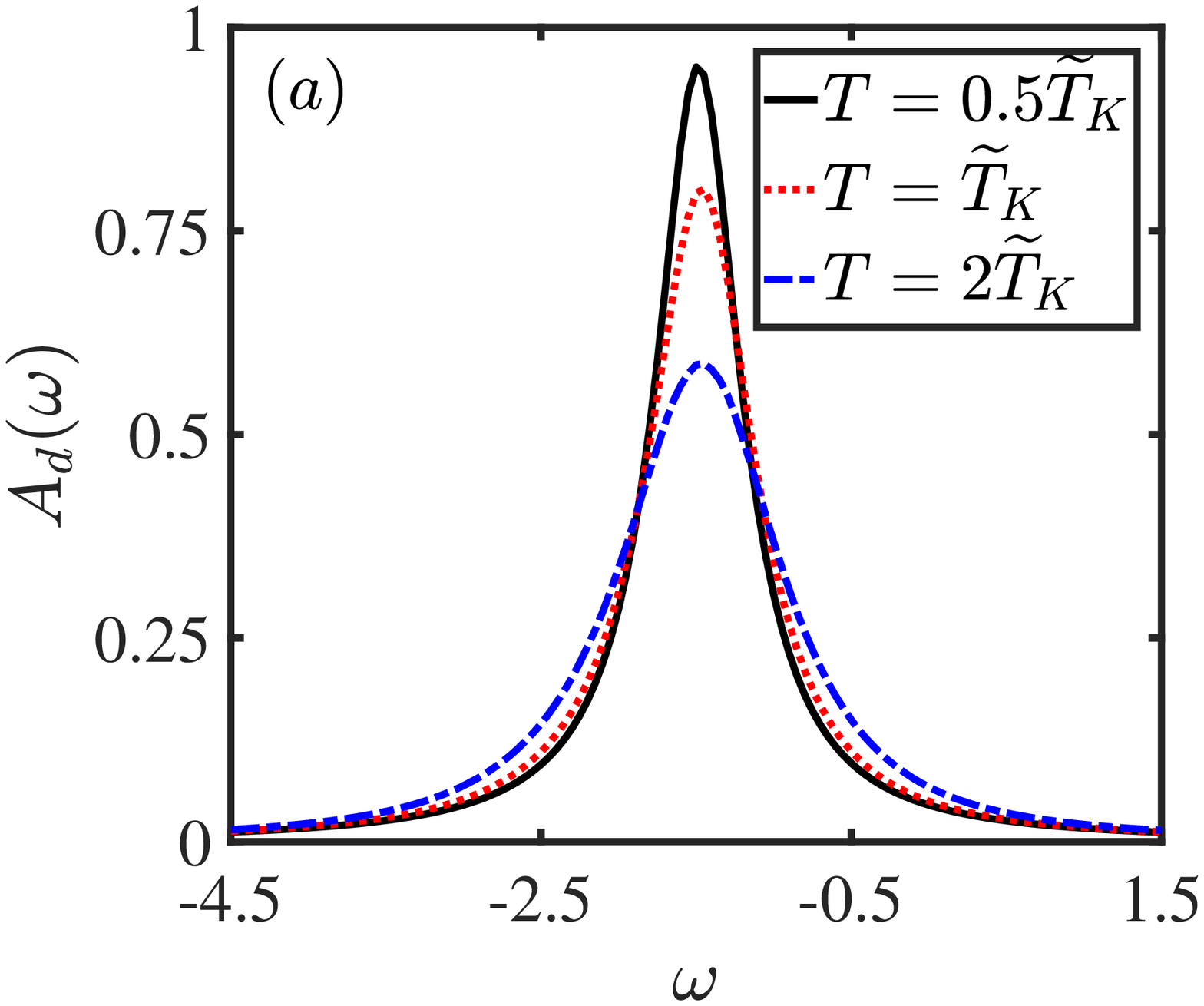}
  \includegraphics[width=0.48\columnwidth]{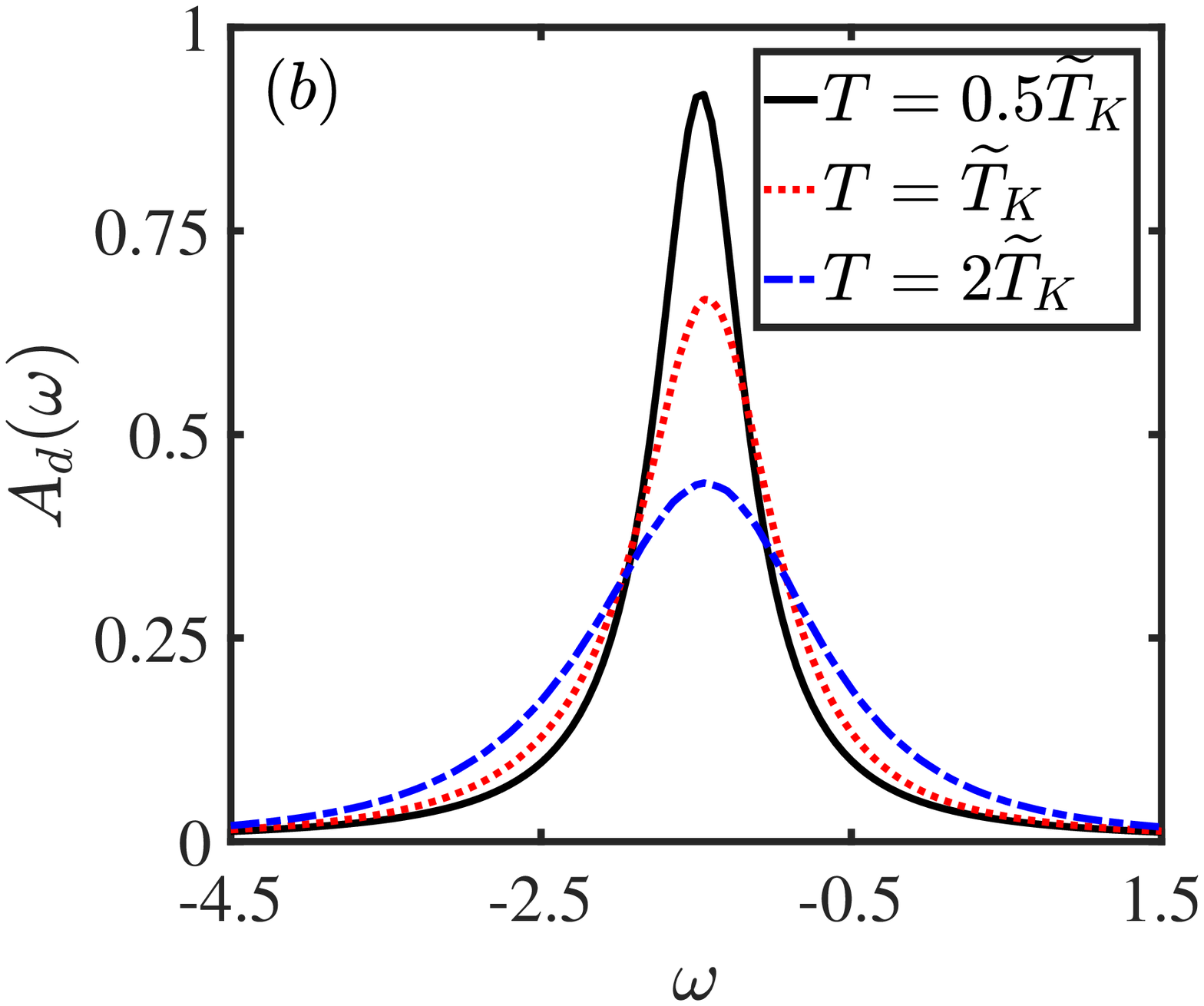}\\
  \vspace{1ex}\vspace{1ex}
  \includegraphics[width=0.48\columnwidth]{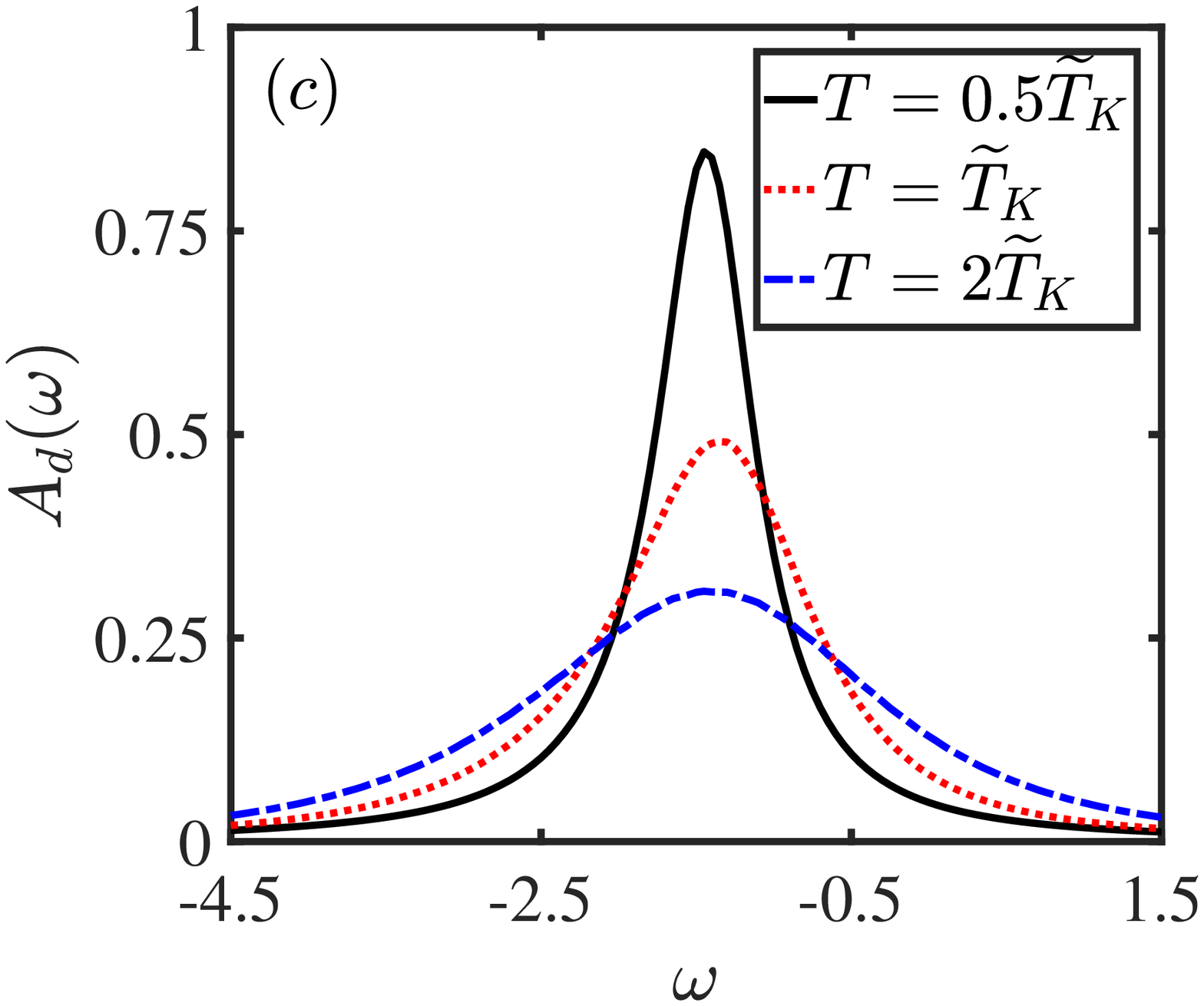}
  \includegraphics[width=0.48\columnwidth]{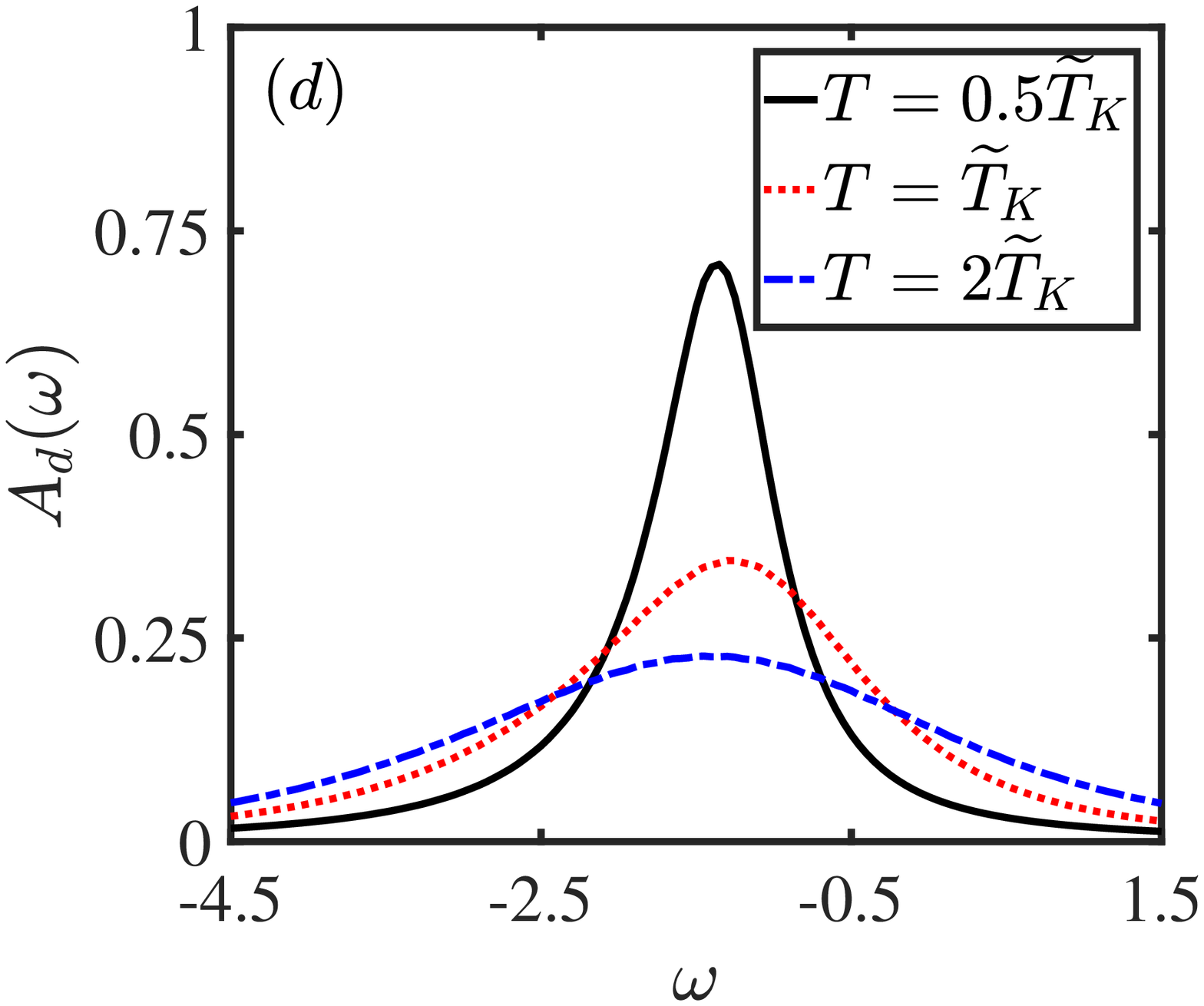}\\
  \vspace{1ex}\vspace{1ex}
  \includegraphics[width=0.48\columnwidth]{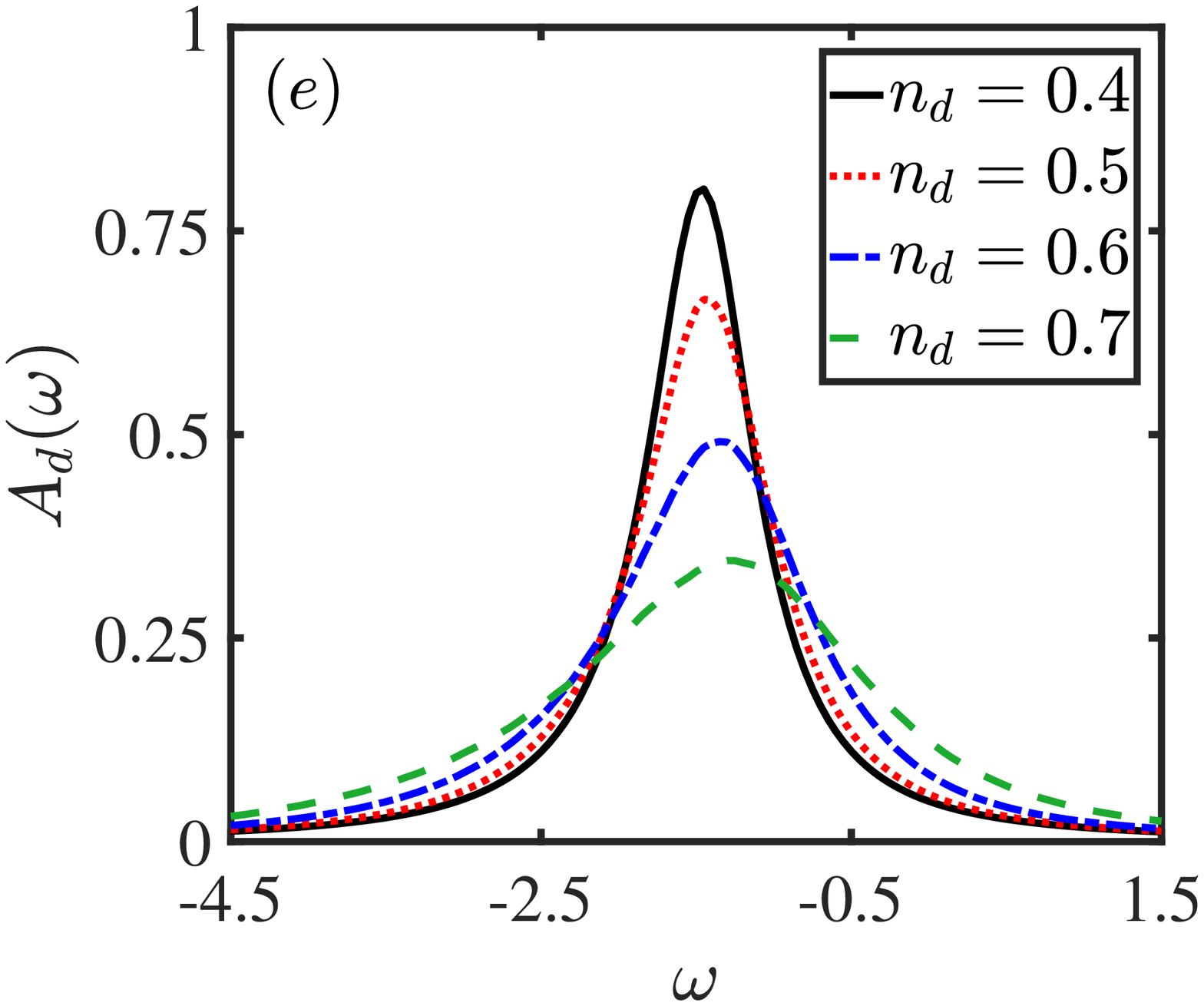}
  \includegraphics[width=0.48\columnwidth]{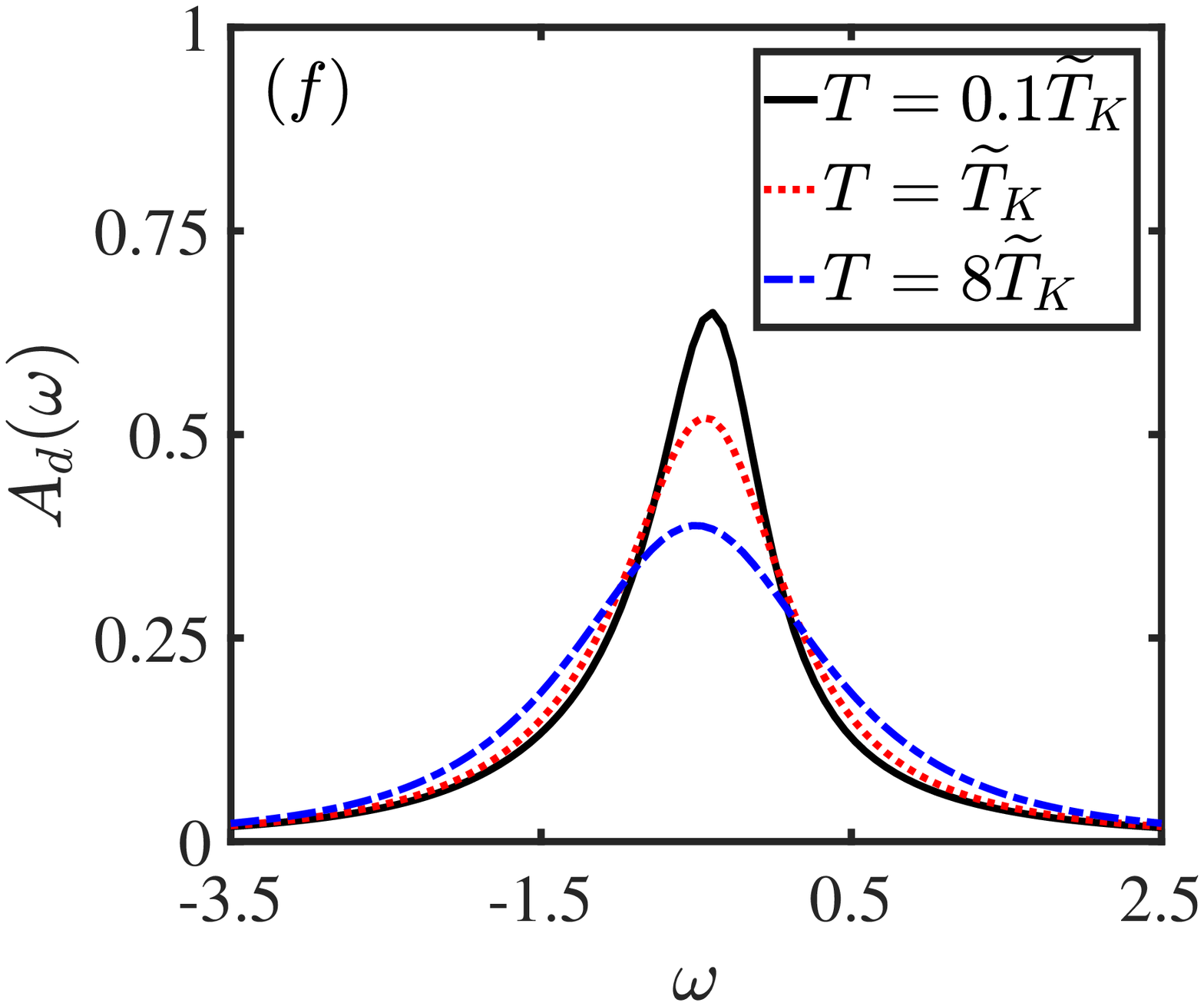}\\
  \caption{The impurity electron DOS $A_{d}(\omega)$ as a function of the frequency $\omega$. Subplots (a)-(d) are for different temperature $T$ with the limited impurity electron concentrations $n_{d}=0.4, 0.5, 0.6, 0.7$ respectively. Subplot (e) describes $A_{d}(\omega)$ become flat versus $n_{d}$ at $T=\widetilde{T}_{K}=1.2387$. In subplot (f), all lines are plotted with $V=1, U=1, D=4, E_{d}=-U/2$ at $n_{d}=0.7$, and $\widetilde{T}_{K}=0.3342$, where the black dotted line is the DOS of FL. The parameters of subplots (a)-(e) are the same as Fig. \ref{fig:p1}.}\label{fig:p6}
\end{figure}

The DOS of the impurity electron is
\begin{eqnarray}\label{eq:19}
A_{d}(\omega)&&=-\frac{1}{\pi}\textrm{Im}G_{d}(\omega+i\delta)\nonumber\\
&&=-\frac{1}{\pi}\textrm{Im}\left[\frac{1}{\omega+i\delta-E_{d}+i\Delta-\Sigma(\omega+i\delta)}\right],
\end{eqnarray}
which is shown in Fig.~\ref{fig:p6}. At the fixed impurity electron concentrations, peaks of the impurity electron DOS decrease versus temperature, and lines of $A_{d}(\omega)$ become flat as shown in Fig.~\ref{fig:p6} (a)-(d). When $T=\widetilde{T}_{K}$, peaks of $A_{d}(\omega)$ decreases versus $n_{d}$ as shown in Fig.~\ref{fig:p6} (e). In Fig.~\ref{fig:p6} (f), the black dotted line describes the DOS of FL, both lines are similar to Fig.~\ref{fig:p6} (a)-(d). The hybridization strength $V$ affects on the Lorentz lineshape of the impurity electron DOS, which induces the scattering between conduction electrons and the impurity electron. This behavior is similar to Kondo physics, but SYKAM does not have the sharp-peak like Kondo resonance around the Fermi surface. \cite{Hewson1997,Coleman2015}

According to Eq.~(\ref{eq:14}), the system behaves as a FL at low temperature, while at high temperature it is in a NFL. Due to Ref.~[\onlinecite{PhysRevB.95.205105}, \onlinecite{PhysRevB.95.134302}], there may exist a quantum phase transition in our model. We have got its entropy $S_{d}$ and specific heat capacity $C_{\textrm{v}}$ versus temperature $T$ by Eq.~(\ref{eq:11}), which is shown in Fig.~\ref{fig:p1}. For intermediate-$T$, $S_{d}$ increases with larger $n_{d}$. All $S_{d}$ gradually approach saturation at high-$T$ limit with $\lim_{T\rightarrow\infty}S_{d}=\ln 2$, as shown in Fig.~\ref{fig:p1} (a).\cite{PhysRevB.94.035135,PhysRevX.7.031006} $C_{\textrm{v}}$ has a maximum at intermediate-$T$, which is similar to the SYK model.\cite{PhysRevX.7.031006,PTEP.2017.083I01} $S_{d}$ increases with larger $n_{d}$ at low temperature, while it decreases for high temperature.

Since entropy and specific heat capacity are continuous and smooth, we expect a crossover, instead of phase transition, between FL and NFL. Interestingly, similar transport properties also display in Kondo physics, and dilute magnetic alloy systems undergo a crossover from free local moment state to the non-magnetic FL state.\cite{Coleman2015,Hewson1997}

However, the difference between Kondo effect and SYKAM is that SYKAM has the NFL behavior $\rho_{c}(T)\propto T^{-1/2}$ below $\widetilde{T}_{K}$ (the resistivity minimum point), while the Kondo screen is the localized FL. Two physical systems have similar transport and thermodynamics, and both undergo a crossover rather than phase transition.\cite{Hewson1997} The reason is that both systems have the scattering between conduction electrons and the impurity electron, but the DOS of the impurity electron is different. Since no symmetry-breaking is involved, a crossover is expected, however thermodynamics is unable to detect it due to the lack of singularity.

To proceed, we apply the RG theory. We begin with the effective action Eq. (\ref{eq:16}) by Fourier transformations, $d^{\dag}_{i}(\tau)=\frac{1}{\sqrt{\beta}}\sum_{n}e^{-i\omega_{n}\tau}d^{\dag}_{i}(i\omega_{n})$ and $d_{i}(\tau^{\prime})=\frac{1}{\sqrt{\beta}}\sum_{n}e^{i\omega_{n}\tau^{\prime}}d_{i}(i\omega_{n})$. We can get
\begin{eqnarray}
\mathcal{S}_{d}&&\simeq-\frac{U^{2}}{4N^{3}(\sqrt{\beta})^{8}}\int d\tau\int d\tau^{\prime}\sum_{ijml}\sum^{8}_{ns,s=1}\left\{d^{\dag}_{i}(i\omega_{n1})d_{i}(i\omega_{n2})\right.\nonumber\\
&&\quad d^{\dag}_{j}(i\omega_{n3})d_{j}(i\omega_{n4})d^{\dag}_{m}(i\omega_{n5})d_{m}(i\omega_{n6})d^{\dag}_{l}(i\omega_{n7})d_{l}(i\omega_{n8})\nonumber\\
&&\quad\left.e^{[-i(\omega_{n1}+\omega_{n3}-\omega_{n6}-\omega_{n8})\tau+i(\omega_{n2}+\omega_{n4}-\omega_{n5}-\omega_{n7})\tau^{\prime}]}\right\}\nonumber\\
&&=-\frac{U^{2}N}{4N^{3}(\sqrt{\beta})^{8}}\sum^{8}_{ns,s=1}\left[d^{\dag}_{i}(i\omega_{n1})d_{i}(i\omega_{n2})d^{\dag}_{j}(i\omega_{n3})d_{j}(i\omega_{n4})\right.\nonumber\\
&&\quad d^{\dag}_{m}(i\omega_{n5})d_{m}(i\omega_{n6})d^{\dag}_{l}(i\omega_{n7})d_{l}(i\omega_{n8})\delta\left(\omega_{n1}-\omega_{n2}+\omega_{n3}\right.\nonumber\\
&&\quad-\omega_{n4}+\omega_{n5}-\omega_{n6}+\omega_{n7}-\omega_{n8})\Big].
\end{eqnarray}
In order to renormalize our model, we have a parameter $\lambda$ ($\lambda>1$) to cut off the fermionic Matsubra frequency $\omega_{n}$ and divide our model into the low energy part and the high energy part as follows
\begin{eqnarray}
&&d^{\dag}_i(\tau)=\frac{1}{\sqrt{\beta}}\sum_{n}e^{-i\omega_{n}\tau}\left(d^{\dag}_{iL}(i\omega_{n})+d^{\dag}_{iH}(i\omega_{n})\right),\\
&&d_{i}(\tau)=\frac{1}{\sqrt{\beta}}\sum_{n}e^{i\omega_{n}\tau}\Big(d_{iL}(i\omega_{n})+d_{iH}(i\omega_{n})\Big),
\end{eqnarray}
where $d^{\dag}_{iL}(i\omega_{n})$ ($d_{iL}(i\omega_{n})$) denotes the low energy component of system, $d^{\dag}_{iH}(i\omega_{n})$ ($d_{iH}(i\omega_{n})$) is the system's high energy component. On the basis of the analytic continuation, it is alluded to a cutoff energy $\frac{\omega_{F}}{\lambda}$ ($\lambda>1$), the system is divided into the low energy component ($0<|\omega|<\frac{\omega_{F}}{\lambda}$) and the high energy component ($\frac{\omega_{F}}{\lambda}<|\omega|<\omega_{F}$), and $\omega_{F}$ is Fermi energy. We rescale the energy as
\begin{eqnarray}
\omega^{\prime}=\lambda\omega\quad(\lambda>1).
\end{eqnarray}
The effective action is given by
\begin{equation}
\mathcal{S}_{int}\approx \mathcal{S}_{L}+\mathcal{S}_{H}+\mathcal{V}(L,H),
\end{equation}
where $\mathcal{S}_{L}$ defines the low energy part of the effective action, $\mathcal{S}_{H}$ is the effective action for high energy component, and $\mathcal{V}(L,H)$ describes the coupling of the low and high energy components.
Hence, the partition function of the system is $\mathcal{Z}=\mathcal{Z}_{c}\mathcal{Z}_{f}$, and $\mathcal{Z}_{f}$ can be read as
\begin{eqnarray}
\mathcal{Z}_{f}&&\approx\left[\int\mathcal{D}d\prod^{8}_{s=1}\int^{\frac{\omega_{F}}{\lambda}}_{0} d\omega_{s}e^{\mathcal{S}_{L}}\prod^{8}_{s=1}\int^{\omega_{F}}_{\frac{\omega_{F}}{\lambda}}d\omega_{s}e^{\mathcal{S}_{H}}\right.\nonumber\\
&&\quad\frac{\prod^{8}_{s=1}\int^{\omega_{F}}_{\frac{\omega_{F}}{\lambda}} d\omega_{s}e^{\mathcal{S}_{H}}e^{\mathcal{V}(L,H)}}{\prod^{8}_{s=1}\int^{\omega_{F}}_{\frac{\omega_{F}}{\lambda}}d\omega_{s}e^{\mathcal{S}_{H}}}\Bigg]\nonumber\\
&&=\int\mathcal{D}d\prod^{8}_{s=1}\int^{\frac{\omega_{F}}{\lambda}}_{0} d\omega_{s}e^{\mathcal{S}_{L}}\prod^{8}_{s=1}\int^{\omega_{F}}_{\frac{\omega_{F}}{\lambda}}d\omega_{s}e^{\mathcal{S}_{H}}e^{\langle\mathcal{V}\rangle},
\end{eqnarray}
where
\begin{eqnarray}
&&e^{\langle\mathcal{V}\rangle}=e^{\langle\mathcal{V}\rangle_{0}+\frac{1}{2}\langle\mathcal{V}^{2}\rangle_{0}+\mathcal{O}(\mathcal{V}^{2})},\\
&&\langle\mathcal{V}\rangle_{0}=\frac{\prod^{8}_{s=1}\int^{\omega_{F}}_{\frac{\omega_{F}}{\lambda}} d\omega_{s}e^{\mathcal{S}_{H}}\mathcal{V}}{\prod^{8}_{s=1}\int^{\omega_{F}}_{\frac{\omega_{F}}{\lambda}}d\omega_{s}e^{\mathcal{S}_{H}}}.
\end{eqnarray}

\begin{figure}
\centering
\includegraphics[width=0.9\columnwidth]{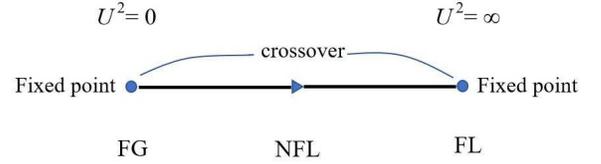}
\caption{Schematic illustration of RG flow in the SYKAM. For $U^{2}=0$, it scales to a weak coupling fixed point, forming the FG. $U^{2}=\infty$ is a strong coupling fixed point of FL. Scaling proceeds from a repulsive fixed point via a crossover to an attractive fixed point, in which exists a NFL at finite $U^{2}$. }\label{fig:p5}
\end{figure}

In the light of the first-order and second-order correction, the RG transformation relation can be written as
\begin{equation}\label{eq:17}
\left(U^{\prime}\right)^{2}=U^{2}\left[1+\frac{U^{2}\omega^{4}_{F}}{2\Delta^{4}}\left(1-\frac{1}{\lambda}\right)^{4}\right],
\end{equation}
we set $\lambda=e^{l}$, and its flow equation is
\begin{equation}\label{eq:18}
\frac{dU^{2}}{dl}=\frac{2U^{4}\omega^{4}_{F}}{\Delta^{4}}.
\end{equation}
We assume $\frac{dU^{2}}{dl}=0$, it demonstrates that SYKAM has two fixed points $U^{2}=0$ and $U^{2}=\infty$ as shown in Fig. \ref{fig:p5}. For $U^{2}=0$, it scales to a weak coupling repulsive fixed point, forming FG. $U^{2}=\infty$ is a strong coupling attractive fixed point of FL. Scaling proceeds from a repulsive fixed point via a crossover to an attractive fixed point in which exists NFL at finite $U^{2}$.

\section{SUMMARY}\label{sec:7}
We have computed the transport and thermodynamics of SYKAM, which demonstrates that it exists a NFL at the lower temperature $T^{\star}<T<\widetilde{T}_{K}$. The RG analysis shows a crossover between FG ($U^{2}=0$) and FL ($U^{2}=\infty$), in which presents NFL at finite $U^{2}$. The impurity electron's entropy $S_{d}$ and the specific heat capacity $C_{\textrm{v}}$ present the similarity to SYK model and Kondo system.\cite{Hewson1997,PhysRevX.7.031006,PTEP.2017.083I01} The resistivity of SYKAM has a minimum at the temperature $\widetilde{T}_{K}$, similar to Kondo temperature, but the impurity electron DOS of SYKAM does not have the sharp Kondo resonance peak around Fermi surface.\cite{Hewson1997,Coleman2015}.

Our model is an extension of single-impurity Anderson model and SYK model. Comparing with the single-impurity Anderson model, our system does not form local moments because the SYK random interaction can not provide the localized interaction to the onsite different impurity spin states. Compared by SYK model, our system has the hybridization between the conduction electrons and impurity electron, which presents at the Kondo systems and does not exist in SYK model. The resistivity minimum emerges in SYKAM when it behaves like a NFL, where the impurity electron has the SYK interaction without Coulomb interaction. Anderson model and our model both have the scattering between the conduction electrons and the impurity electron because of the hybridization, and except that the DOS of the impurity electron does not display the sharp-peak as the Kondo resonance and not have the localized magnetic moments in SYKAM.

Finally, it may be helpful to lead to a new route to realize the Kondo systems (heavy-fermion compounds ,various quantum dot devices and novel material Kondo systems) and the random interaction SYK physics. With the development of ultracold atom technique for realizing Kondo lattice model,\cite{RevModPhys.82.1225,PhysRevA.84.053619,PhysRevLett.115.165302} our results may be protocolled in near-future experiments.
\section*{ACKNOWLEDGMENTS}
This research was supported in part by NSFC under Grant No. 11674139, No. 11704166, No. 11834005, the Fundamental Research Funds for the Central Universities, and PCSIRT (Grant No. IRT-16R35).
\bibliographystyle{apsrev4-1}
\bibliography{SYK}

\end{document}